\documentclass[12pt]{article}
\usepackage{amsmath,amssymb,amsfonts}
\usepackage{graphicx}
\def\br(#1,#2){\left\langle#1#2\right\rangle}
\def\sq(#1,#2){\left[#1#2\right]}
\def\s(#1,#2){s_{#1 #2}}
\def\t(#1,#2,#3){s_{#1 #2 #3}}

\begin{document}

\begin{titlepage}
\hspace*{\fill}\parbox[t]{3.5cm}
{
\today} \vskip2cm
\begin{center}
{\Large \bf Higgs Decay to Two Photons} \\\bigskip

\bigskip\bigskip\bigskip\bigskip

{\bf William J.\ Marciano,$^1$ Cen Zhang,$^2$ and Scott Willenbrock$\,^2$}\\
\bigskip\bigskip\bigskip
$^1$ Physics Department, Brookhaven National Laboratory, Upton, NY  11973 \\
\bigskip
$^2$ Department of Physics, University of Illinois at Urbana-Champaign \\ 1110 West Green Street, Urbana, IL  61801
\end{center}

\bigskip\bigskip\bigskip\bigskip

\begin{abstract}
The amplitude for Higgs decay to two photons is calculated in renormalizable and unitary gauges using dimensional regularization at intermediate steps.  The result is finite, gauge independent, and in agreement with previously published results. The large Higgs mass limit is examined using the Goldstone-boson
equivalence theorem as a check on the use of dimensional regularization and to explain
the absence of decoupling.
\end{abstract}

\end{titlepage}

\section{Introduction}\label{sec:intro}

One of the primary ways to search for the Higgs boson at the CERN Large Hadron Collider is via its decay to two photons.
That decay is induced by quantum loop corrections involving the $W$ boson and fermions, primarily the top quark. The gauge invariant decay amplitude is given by
\begin{equation}
{\cal M}=\frac{e^2g}{(4\pi)^2m_W}F
(k_1\cdot k_2g^{\mu\nu}-k_2^\mu k_1^\nu)\epsilon_\mu(k_1) \epsilon_\nu(k_2)
\end{equation}
where $F$ includes contributions from $W$ loops and fermion loops:
\begin{equation}
F=F_W(\beta_W)+\sum_fN_cQ_f^2F_f(\beta_f)
\end{equation}
and $N_c$ is a color factor ($N_c =1$ for leptons, $N_c=3$ for quarks),
\begin{equation}
\beta_W=\frac{4m_W^2}{m_H^2},\quad
\beta_f=\frac{4m_f^2}{m_H^2}.
\end{equation}
with
\begin{eqnarray}
F_W(\beta)&=&2+3\beta+3\beta(2-\beta)f(\beta)\label{eq:standard}
\\
F_f(\beta)&=&-2\beta\left[1+(1-\beta)f(\beta)\right]\label{eq:fermion}
\end{eqnarray}
where
\begin{equation}
f(\beta)=\left\{
\begin{array}{ll}
\arcsin^2(\beta^{-\frac{1}{2}}) &\quad \mbox{for}\quad \beta\geq 1\\
-\frac{1}{4}\left[\ln\frac{1+\sqrt{1-\beta}}{1-\sqrt{1-\beta}}-i\pi\right]^2&
\quad \mbox{for}\quad \beta<1
\end{array}\right.\;.
\end{equation}

The value of $F_W$ in the limit of small Higgs mass was first calculated by
Ellis {\it et al.} \cite{Ellis:1975ap} and found to have a numerical value of 7. The general result for
arbitrary $m_H$ was later calculated by Shifman {\it et al.}
in 't Hooft-Feynman linear and non-linear gauges \cite{Shifman:1979eb}.
This result agrees with Ref.~\cite{Ellis:1975ap} in the small Higgs mass limit.
It leads to the predicted decay rate:
\begin{equation}
\Gamma(H\to\gamma\gamma)=|F|^2\left(\frac{\alpha}{4\pi}\right)^2\frac{G_Fm_H^3}{8\sqrt{2}\pi}
\end{equation}
That prediction not only tests the Standard Model, but also provides a test of additional ``New Physics'' effects that might contribute at the loop level and modify the decay rate.

A recent pair of papers \cite{Gastmans:2011ks,Gastmans:2011wh} has questioned the correctness of the $W$ loop contribution to $H\to \gamma\gamma$ and the validity of dimensional regularization.  These papers disagree with the results in Refs.~\cite{Ellis:1975ap,Shifman:1979eb}.  For this reason it is timely to revisit the calculation of this process and to settle the issue of the correct expression for the decay amplitude. We also discuss the use of dimensional regularization, and the behavior of the amplitude in the limit that the Higgs boson is much heavier than the loop particles.

\section{Unitary gauge}

\begin{figure}[htb]
\centering\includegraphics[width=14cm]{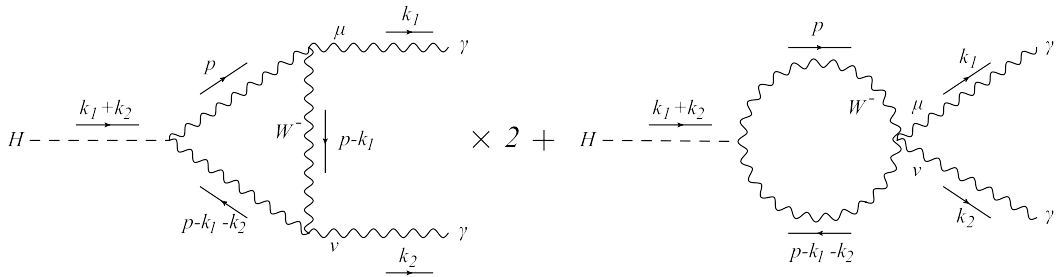}
\caption{Feynman diagrams for $H\to \gamma\gamma$ in unitary gauge.}\label{fig:unitary}
\end{figure}

We begin by calculating the diagrams in Fig.~\ref{fig:unitary} in unitary gauge \cite{Weinberg:1971fb} using dimensional regularization \cite{'tHooft:1972fi}.  To the best of our knowledge, this is a new calculation; we present the details in Appendix A.  Although the unitary gauge is often avoided in $W$ boson loop calculations because of the
large amount of algebra and high degree of ultraviolet divergences encountered, the use of modern computing algorithms and dimensional regularization make such calculations relatively straightforward.  The advantage of the unitary gauge is that it involves only physical particles and avoids ghost and Goldstone boson loops.
Hence the number of Feynman diagrams is minimal.
We find the standard result (given in Eq.~(\ref{eq:standard})),
\begin{equation}
{\cal M}=\frac{e^2g}{(4\pi)^2m_W}\left[2+3\beta+3\beta(2-\beta)f(\beta)\right]
(k_1\cdot k_2g^{\mu\nu}-k_2^\mu k_1^\nu)\epsilon_\mu(k_1) \epsilon_\nu(k_2)
\label{eq:amplitude}\end{equation}
where $\beta =4m_W^2/m_H^2$. This result agrees with Refs.~\cite{Ellis:1975ap,Shifman:1979eb}.
The use of dimensional regularization ensures a result that respects electromagnetic gauge invariance; that is, the amplitude vanishes if either photon polarization vector is replaced by its four-momentum.

\section{Renormalizable gauge}

\begin{figure}[htb]
\centering\includegraphics[width=14cm]{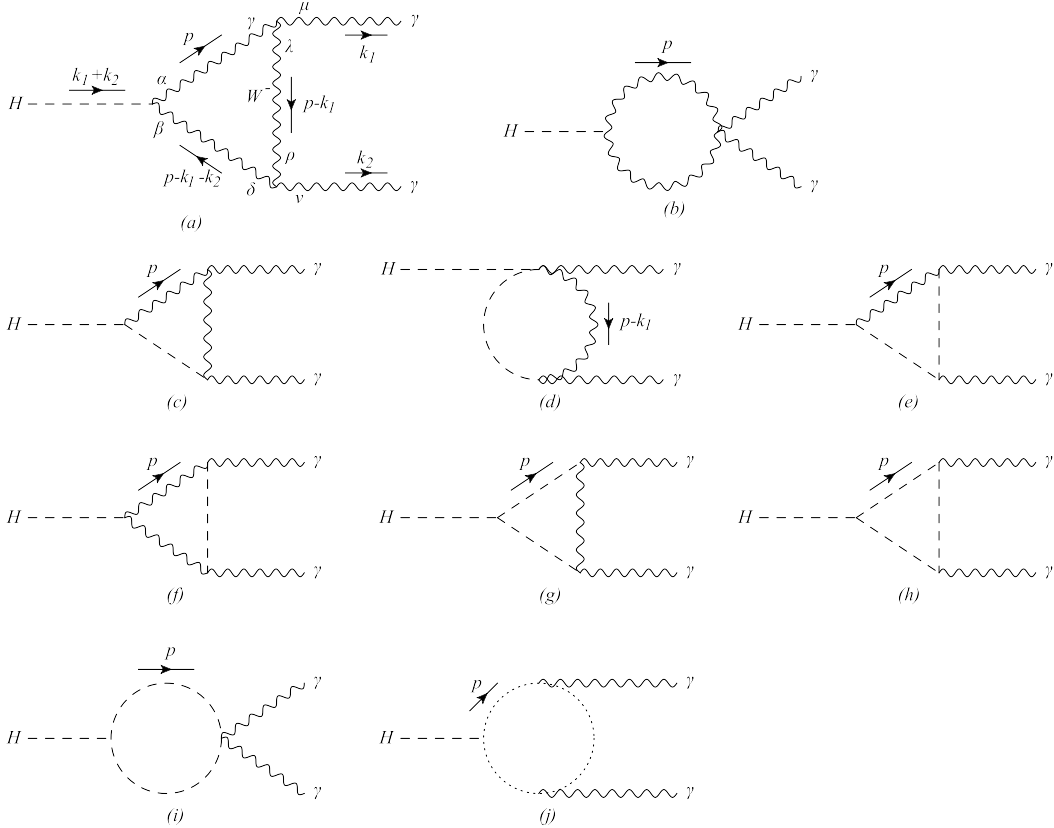}
\caption{Feynman diagrams for $H\to \gamma\gamma$ in the $R_\xi$ gauge. Diagrams that can be obtained by exchanging the two photons and by charge conjugation are omitted. Instead, we include a factor of 2 in diagrams (a,f,g,h) and a factor of 4 in diagrams (c,d,e,j) to include the contributions from these diagrams. Diagram (j) also contains a factor of $-1$ from the ghost loop.\label{fig:Rxi}}
\end{figure}

We repeat the calculation, this time in renormalizable gauges \cite{Fujikawa:1972fe} ($R_\xi$ gauge, for arbitrary $\xi$), again using dimensional regularization.  There are many more diagrams than in unitary gauge, as shown in Fig.~\ref{fig:Rxi}.  To the best of our knowledge, this calculation has been performed previously only in 't Hooft-Feynman gauge ($\xi=1$), first in Ref.~\cite{Ellis:1975ap} in the limit of a light Higgs boson (see also Ref.~\cite{Gavela:1981ri} for non-linear gauge), and later in Ref.~\cite{Shifman:1979eb} for arbitrary Higgs mass.

We present the details in Appendix B.  The contribution from loops of Goldstone bosons is
\begin{equation}
{\cal M}_{GB}=
\frac{e^2g}{(4\pi)^2m_W}\left[2-8\frac{\xi m_W^2}{m_H^2}f\left(\frac{4 \xi m_W^2}{m_H^2}\right)\right](k_1\cdot k_2g^{\mu\nu}-k_2^\mu k_1^\nu)\epsilon_\mu(k_1) \epsilon_\nu(k_2)\label{eq:Goldstone}
\end{equation}
The $\xi$-dependent term is cancelled by contributions from other diagrams.  The first term in the brackets is responsible for the first term in Eq.~(\ref{eq:amplitude}) of the complete result.  This term, which survives in the limit $\beta\to 0$, will be discussed in a later section.  The remaining diagrams in Fig.~\ref{fig:Rxi}, together with the Goldstone boson loop diagrams, yield the standard result given in Eq.~(\ref{eq:amplitude}).

\section{Dimensional Regularization}\label{sec:dimreg}

Our calculations confirm the standard result for the amplitude for Higgs decay to two photons via a $W$ boson loop given in Eq.~(\ref{eq:amplitude}).  The difference between the standard result and the result in Refs.~\cite{Gastmans:2011ks,Gastmans:2011wh} is traced back to a certain integral,
\begin{equation}
I_{\mu\nu}(n)=\int d^n\ell \frac{\ell^2g_{\mu\nu}-4\ell_\mu\ell_\nu}{(\ell^2-M^2+i\epsilon)^3}\;.
\end{equation}
Refs.~\cite{Gastmans:2011ks,Gastmans:2011wh} shun dimensional regularization, and therefore set $n=4$ throughout their calculation.  They argue that symmetric integration implies that the numerator of the above integral vanishes, and, therefore, conclude that $I_{\mu\nu}(4)=0$.

The flaw in this argument is that there is a cancellation between two integrals, each of which is ultraviolet divergent, so a regulator is needed to make sense of the calculation.  Dimensional regularization provides a regulator which respects gauge invariance \cite{'tHooft:1972fi}, so it is ideal for such a calculation.  Evaluating the integral using dimensional regularization yields
\begin{equation}
I_{\mu\nu}(n)=-\frac{i\pi^{n/2}}{2}\Gamma\left(3-\frac{n}{2}\right)
\left(\frac{1}{M^2}\right)^{2-n/2}g_{\mu\nu}
\label{eq:I}\end{equation}
which is finite and unambiguous for $n=4$,
\begin{equation}
I_{\mu\nu}(4)=-\frac{i\pi^2}{2}g_{\mu\nu}
\end{equation}
Refs.~\cite{Gastmans:2011ks,Gastmans:2011wh} state that the integral is defined only for $n<4$, but Eq.~(\ref{eq:I}) shows that it is defined in the neighborhood of $n=4$.
There are poles at $n=6,8,...$, but the integral is finite and unambiguous in the neighborhood of $n=4$.

It is incorrect to set this integral to zero for $n=4$ as is done in Refs.~\cite{Gastmans:2011ks,Gastmans:2011wh}.  This is the reason electromagnetic gauge invariance is lost at intermediate steps in that calculation.

\section{Decoupling}

The standard electroweak theory has only one fundamental mass scale, the Higgs vacuum expectation value $v\approx 246$ GeV (or, equivalently, $G_F = 1/(\sqrt{2}v^2)$).  In addition, there are a variety of dimensionless couplings, such as the weak gauge coupling $g$, Yukawa couplings $y$, and the Higgs self-interaction $\lambda$.  The various particles acquire their mass via their coupling to the Higgs vacuum expectation value: $m_W\sim gv$, $m_f \sim yv$, $m_H\sim \sqrt\lambda v$. An amplitude may vanish if one of these dimensionless couplings is set to zero.  This is sometimes called decoupling, although decoupling has another, deeper meaning in quantum field theory, as we discuss below.

\begin{figure}[htb]
\centering\includegraphics[width=9cm]{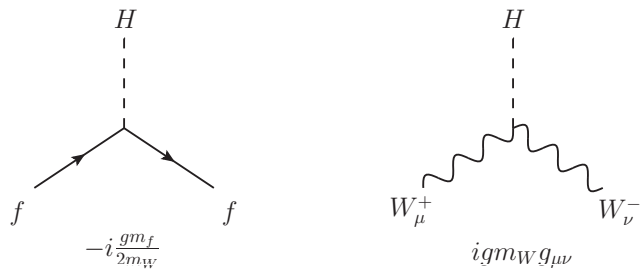}
\caption{Feynman rules for the Higgs coupling to fermions and $W$ bosons.}\label{fig:fr}
\end{figure}

For example, consider the amplitude for the Higgs boson to decay to a pair of heavy fermions at tree level via the coupling shown in Fig.~\ref{fig:fr}.  The amplitude for this process is proportional to $ym_f$.  Consider the limit of a Higgs boson much heavier than the fermion, $m_f/m_H\to 0$.  This limit corresponds to $y/\sqrt\lambda\to 0$, that is, the limit of vanishing Yukawa coupling.  The decay amplitude clearly vanishes in this limit.  The decay of a Higgs to two photons via a heavy fermion loop inherits this decoupling behavior from the tree-level process, as evidenced by the vanishing of Eq.~(\ref{eq:fermion}) in the limit $\beta\to 0$.

In contrast, consider the amplitude for the Higgs boson to decay to a pair of $W$ bosons via the coupling shown in Fig.~\ref{fig:fr}.  Consider the limit of a Higgs boson much heavier than the $W$ boson, $m_W/m_H \sim g/\sqrt\lambda\to 0$.  This is the limit of vanishing gauge coupling.  One might expect the Higgs to decouple from the $W$ bosons in this limit.  However, the polarization vector of a longitudinal (zero helicity) $W$ boson in this limit is approximately $\epsilon^\mu(p)\sim p^\mu/m_W$, so the amplitude for Higgs decay to a pair of longitudinal $W$ bosons is proportional to $gm_W p_1 \cdot p_2/m_W^2 \sim m_H^2/v$.  This amplitude does not decouple.  Hence the decay of a Higgs to two photons via a $W$ loop also does not decouple, as evidenced by the first term in brackets in Eq.~(\ref{eq:amplitude}).

The discussion above has nothing to do with the Appelquist-Carazzone decoupling theorem \cite{Appelquist:1974tg}, which is deeper than taking the limit of vanishing coupling.  This theorem states that the effects of heavy particles on light particles is contained in an unobservable renormalization of the light-particle couplings, plus observable effects that decrease like an inverse power of the heavy particle mass.  The heavy particle does not have vanishing coupling to the light particles as its mass is taken to infinity.  Refs.~\cite{Gastmans:2011ks,Gastmans:2011wh} misconstrue the Appelquist-Carazzone decoupling theorem.

\section{Goldstone-Boson Equivalence Theorem}

A nice check of the large Higgs mass limit of the
$W$ loop contribution to $H\rightarrow\gamma\gamma$ is provided by the
Goldstone-boson equivalence theorem \cite{Cornwall:1974km,Vayonakis:1976vz,Chanowitz:1985hj}.
This theorem states that at high energies ($s\gg m_W^2$) $S$-matrix amplitudes involving
external longitudinal components of $W^{\pm}$ and $Z$ bosons are
equivalent up to $\mathcal{O}(m_W/\sqrt{s})$ to the corresponding amplitudes
in the Higgs-Goldstone scalar theory with Goldstone bosons
replacing $W_L^\pm$ and $Z_L$. A nice feature of replacing longitudinal
gauge bosons with scalar Goldstone bosons is the ease of calculations.

Application of the Goldstone boson equivalence theorem
to quantum loops was initiated in Ref.~\cite{Marciano:1987un} where the
radiative corrections to $H\rightarrow W^+W^-$ and $H\rightarrow ZZ$ were computed
in the large Higgs mass limit employing the Higgs-Goldstone
scalar theory in the Landau gauge. In that gauge,
gauge boson-scalar mixing is avoided and the Goldstone
boson propagators have zero mass.

\begin{figure}[htb]
\centering
\includegraphics[width=10cm]{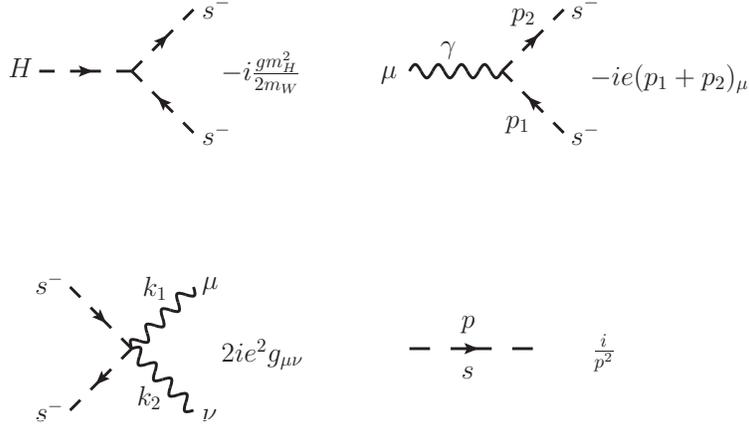}
\caption{Feynman rules for Higgs and Goldstone bosons in Landau gauge.\label{fig:frlandau}}
\end{figure}

For $H\rightarrow\gamma\gamma$, relatively few Feynman
rules are required in the Higgs-Goldstone boson scalar theory.
They are given in Fig.~\ref{fig:frlandau}.

\begin{figure}[htb]
\centering
\includegraphics[width=12cm]{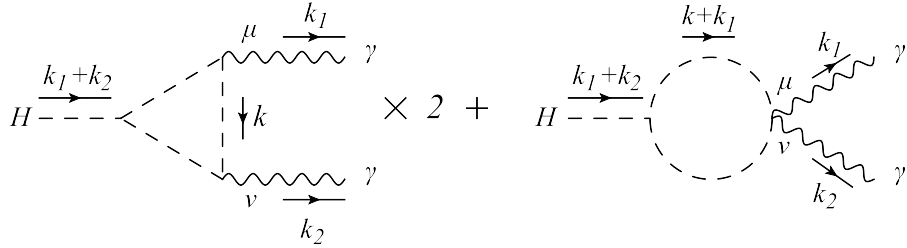}
\caption{Feynman diagrams for $H\to \gamma\gamma$ from Goldstone boson loops.\label{fig:landau}}
\end{figure}

Employing those Feynman rules and combining the amplitudes
in Fig.~\ref{fig:landau}, one finds
\begin{equation}
\label{eq:M1}
{\cal M}_{GB}=i\frac{e^2gm_H^2}{m_W}\epsilon_\mu(k_1)\epsilon_\nu(k_2)
\int\frac{d^4k}{(2\pi)^4}\frac{k^2g^{\mu\nu}-4k^\mu k^\nu}{k^2(k^2+2k\cdot k_1)(k^2-2k\cdot k_2)}\;.
\end{equation}
Although the integral is finite, it contains canceling ultraviolet
divergences. To avoid any ambiguity in the finite part, one must
be certain that electromagnetic gauge invariance is preserved.
To do that requires a regulator, such as dimensional regularization,
which maintains the symmetry.

To proceed further, we combine the propagators in Eq.~(\ref{eq:M1})
using Feynman parameters and obtain
\begin{equation}
{\cal M}_{GB}=ie^2g\frac{m_H^2}{m_W}\epsilon_\mu(k_1)\epsilon_\nu(k_2)
\int^1_02ydy\int^1_0dx\int\frac{d^4\ell}{(2\pi)^4}\frac{N^{\mu\nu}(\ell)}{(\ell^2-\Delta)^3}
\end{equation}
where
\begin{flalign}
\label{eq:M3a}
N^{\mu\nu}(\ell)&=\ell^2g^{\mu\nu}-4\ell^\mu \ell^\nu-2k_1\cdot k_2 y^2x(1-x) g^{\mu\nu}
+4k_2^\mu k_1^\nu y^2x(1-x)\\
\Delta&=-m_H^2y^2x(1-x)
\end{flalign}
At this point, if one applies 4 dimensional symmetric
integration to the first two terms in Eq.~(\ref{eq:M3a}), they appear
to exactly cancel and one is left with an amplitude that is proportional to
$\frac{1}{2}k_1\cdot k_2g^{\mu\nu}-k_1^\nu k_2^\mu$ rather than
$k_1\cdot k_2g^{\mu\nu}-k_1^\nu k_2^\mu$ as required
by electromagnetic gauge invariance. Clearly, the $g^{\mu\nu}$ term
is problematic. One could merely accept the correctness of the
$k_1^\nu k_2^\mu$ coefficient and adjust the $g^{\mu\nu}$ term accordingly. That would
operationally work here, but it would not
in the unitary gauge. There, because of the high degree of divergence
encountered, the coefficient of both the $g^{\mu\nu}$ and $k_1^\nu k_2^\mu$
terms are ambiguous without the use of dimensional regularization.
Consider, for example, the integral ${\cal M}_{1132}$ (and ${\cal M}_{3132}$) in Ref.~\cite{Gastmans:2011wh}, which can be written as
\begin{flalign}
{\cal M}_{1132}&=
\frac{i2e^2g}{m_W}\int\frac{d^4l}{(2\pi)^4}\int_0^1d\alpha_1\int_0^{1-\alpha_1}d\alpha_2
\frac{\ell^2g^{\alpha\beta}-4\ell^\alpha \ell^\beta}{\left[\ell^2-m_W^2+2\alpha_1\alpha_2(k_1\cdot k_2)\right]^3}
\nonumber\\&
\times(k_{1\alpha}g_{\lambda\mu}-k_{1\lambda}g_{\alpha\mu})
(k_{2\beta}{g^\lambda}_{\nu}-k_2^\lambda g_{\beta\nu})
\end{flalign}
This integral contributes to both the $g^{\mu\nu}$ and $k_1^\nu k_2^\mu$
terms, and it is ambiguous without the use of dimensional regularization.  This is one
of the integrals that contributes to the amplitude in the limit $\beta\to 0$.

If instead of using 4 dimensional symmetric
integration in Eq.~(\ref{eq:M3a}), we employ dimensional regularization, with
\begin{equation}
\frac{d^4\ell}{(2\pi)^4}\rightarrow \frac{d^n\ell}{(2\pi)^n}
\end{equation}
and
\begin{equation}
4\ell^\mu \ell^\nu\rightarrow \frac{4}{n}\ell^2g^{\mu\nu},
\end{equation}
we find
\begin{eqnarray}
{\cal M}_{GB}&=&\frac{e^2g}{(4\pi)^2m_W}\epsilon_\mu(k_1)\epsilon_\nu(k_2)
(g^{\mu\nu}k_1\cdot k_2-k_2^\mu k_1^\nu)h(n)
\\
h(n)&=&\frac{4\Gamma^2\left(\frac{n}{2}-1\right)
\Gamma\left(3-\frac{n}{2}\right)}{\Gamma(n-1)}
\left(\frac{-m_H^2}{4\pi}\right)^{\frac{n}{2}-2}
\\
h(4)&=&2
\end{eqnarray}
The amplitude is gauge invariant for all $n$
and continuous at $n=4$. In the $n=4$ and large Higgs mass limit ($\beta\to 0$), it
confirms
the result in Eq.~(\ref{eq:amplitude}) and the non-decoupling of the $W$ loop.  It also confirms the result, given in Eq.~(\ref{eq:Goldstone}), for the contribution from Goldstone boson loops in Laudau gauge ($R_\xi$ gauge for $\xi=0$),
as well as a similar calculation in Ref.~\cite{Korner:1995xd}.
Dimensional regularization played a crucial role in preserving
electromagnetic gauge invariance and
providing a unique Standard Model result.  This is a specific example of the general comments on dimensional regularization made in Section~\ref{sec:dimreg}.

This calculation gives us yet another way to understand the non-decoupling of the $W$ loop contribution to $H\to \gamma\gamma$ in the limit $g\to 0$.  Due to the Higgs coupling to Goldstone bosons shown in Fig.~\ref{fig:frlandau}, the amplitude is proportional to $\lambda v$, which does not vanish in the limit $g\to 0$.

\section{Conclusions}

We have calculated the $W$ boson loop contribution to Higgs decay into two photons in the unitary and renormalizable ($R_\xi$) gauges of the Standard Model.
Using dimensional regularization,  we were able to preserve electromagnetic gauge invariance throughout the calculations and confirm the
classic results of Refs.~\cite{Ellis:1975ap,Shifman:1979eb}.  In so doing, our results can also be viewed as a test of dimensional regularization, a technique that has been
applied to many electroweak and QCD calculations.  Its success here provides a further validation of that important prescription.

Using the Goldstone boson equivalence theorem, we were able to provide an additional check of the large Higgs mass limit of our calculation in
a computationally simple manner.  That approach illustrated how and why a naive interpretation of decoupling fails and further demonstrates the utility of
dimensional regularization in maintaining electromagnetic gauge invariance.

Having confirmed the validity of Refs.~\cite{Ellis:1975ap,Shifman:1979eb} and its unique Standard Model prediction for the Higgs to two photon decay rate, we anxiously await discovery of the Higgs
scalar particle and experimental test of its two photon branching ratio.

\bigskip\bigskip

While this work was being written up,
a preprint by the authors of Ref.~\cite{Shifman:1979eb}
appeared \cite{Shifman:2011ri}. It also criticizes the claims in Refs.\cite{Gastmans:2011ks,Gastmans:2011wh}
and discusses the Goldstone boson equivalence theorem and non-decoupling.
In another preprint \cite{Huang:2011yf}, a different gauge
invariant regulator is used to arrive at the same gauge invariant result as ours.

After this work was submitted for publication,
a preprint by F.~Jegerlehner \cite{Jegerlehner:2011jm} was posted, which reaches conclusions in agreement
with ours.
In addition, we received a
private communication by R.~Jackiw, in which he gives a general
discussion of finite loop ambiguities in quantum field theories and the need to
resolve them by physics input or symmetry considerations \cite{Jackiw:1999qq}. As we have shown, for the $H\to\gamma\gamma$ amplitude under consideration, the requirement of electromagnetic gauge invariance resolves any ambiguity and leads to a unique finite result.

\section*{Acknowledgements}

This material is based upon work
supported in part by the U.~S.~Department of Energy under contracts
No.~DE-FG02-91ER40677 and DE-AC02-76CH00016.

\appendix

\section{Unitary Gauge}

In the unitary gauge, ghosts and Goldstone bosons are absent.
There are two $W$ loop diagrams for the decay,
shown in Fig.~\ref{fig:unitary}.
Another diagram can be obtained by exchanging the two photons
in the first diagram. Since it gives the same amplitude, we simply include a factor
of 2 in the following calculation.

The momenta of the particles are labeled in Fig.~\ref{fig:unitary};
$k_1$ and $k_2$ are the momenta of the photons, so
\begin{equation}
k_1^2=k_2^2=0\;,
\end{equation}
and
\begin{equation}
\epsilon_\mu(k_1)k_1^\mu=\epsilon_\nu(k_2)k_2^\nu=0\;,
\end{equation}
since we are dealing with real photons.
The four momentum of the Higgs particle is $k_1+k_2$, so
\begin{equation}
2(k_1\cdot k_2)=m_H^2
\end{equation}
where $m_H$ is the mass of the Higgs boson.

It is straightforward to write down the amplitude. After some algebra,
the total amplitude is
\begin{equation}
i{\cal M}=\int\frac{d^np}{(2\pi)^n}\left(i{\cal M}_1 g^{\mu\nu}+i{\cal M}_2 p^\mu p^\nu+i{\cal M}_3 p^\mu k_1^\nu+i{\cal M}_4 k_2^\mu p^\nu+i{\cal M}_5 k_2^\mu k_1^\nu\right)\epsilon_\mu(k_1) \epsilon_\nu(k_2)
\end{equation}
where
\begin{flalign}
i{\cal M}_1=&-\frac{2e^2g}{m_W^3}\frac
{1}
{(p^2-m_W^2)[(p-k_1)^2-m_W^2][(p-k_1-k_2)^2-m_W^2]}
\nonumber\\
&\times\left\{
2(p\cdot k_1)^3-2(p\cdot k_1)(p\cdot k_2)^2+2(p^2-3m_W^2)(p\cdot k_1)(p\cdot k_2)
\right.\nonumber\\&\left.
-3(p^2-m_W^2)(p\cdot k_1)^2+(p^2-m_W^2)(p\cdot k_2)^2
+\left[(p^2-m_W^2)^2+2(1-n)m_W^4\right](p\cdot k_1)
\right.\nonumber\\&\left.
-(p^2-m_W^2)^2(p\cdot k_2)
+m_W^2\left[\left((n-1)m_W^2+m_H^2\right)(p^2-m_W^2)+4m_W^2m_H^2\right]
\right\}\\
i{\cal M}_2=&\frac{4e^2g}{m_W}\frac{m_H^2+2(n-1)m_W^2}
{(p^2-m_W^2)[(p-k_1)^2-m_W^2][(p-k_1-k_2)^2-m_W^2]}\\
i{\cal M}_3=&\frac{e^2g}{m_W^3}\frac{1}{(p^2-m_W^2)[(p-k_1)^2-m_W^2][(p-k_1-k_2)^2-m_W^2]}
\nonumber\\&\times
\left[
(p^2)^2-3(p^2-3m_W^2)(p\cdot k_1)-(p^2+7m_W^2)(p\cdot k_2)-5p^2m_W^2
\right.\nonumber\\&\left.
+2(p\cdot k_1)(p\cdot k_2)+2(p\cdot k_1)^2-4(2n-3)m_W^4
\right]\\
i{\cal M}_4=&-\frac{e^2g}{m_W^3}\frac{1}{(p^2-m_W^2)[(p-k_1)^2-m_W^2][(p-k_1-k_2)^2-m_W^2]}
\nonumber\\&\times
\left[
(p^2-m_W^2)(p^2-4m_W^2)-(3p^2-17m_W^2)(p\cdot k_1)-(p^2-m_W^2)(p\cdot k_2)
\right.\nonumber\\&\left.
+2(p\cdot k_1)(p\cdot k_2)+2(p\cdot k_1)^2
\right]\\
i{\cal M}_5=&\frac{4e^2g}{m_W}\frac{p^2+3m_W^2}
{(p^2-m_W^2)[(p-k_1)^2-m_W^2][(p-k_1-k_2)^2-m_W^2]}
\end{flalign}

We rewrite ${\cal M}_1$, ${\cal M}_3$ and ${\cal M}_4$ as:
\begin{flalign}
i{\cal M}_1=&
\frac{2e^2g}{m_W^3}\left[
-\frac{2p\cdot(k_1-k_2)+m_H^2}{4(p^2-m_W^2)}-\frac{m_W^2}{(p-k_1)^2-m_W^2}
-\frac{2p\cdot(k_1-k_2)-m_H^2-4m_W^2}{4\left[(p-k_1-k_2)^2-m_W^2\right]}
\right.\nonumber\\&\left.
-\frac{4(m_H^2+2m_W^2)(p\cdot k_2)-4(1-n)m_W^4-m_H^4}{4(p^2-m_W^2)\left[(p-k_1-k_2)^2-m_W^2\right]}
\right.\nonumber\\&\left.
-\frac{4m_H^2m_W^4}{(p^2-m_W^2)\left[(p-k_1)^2-m_W^2\right]\left[(p-k_1-k_2)^2-m_W^2\right]}
\right]
\\
i{\cal M}_3=&\frac{e^2g}{m_W^3}
\left[
\frac{1}{2(p^2-m_W^2)}+\frac{1}{2\left[(p-k_1-k_2)^2-m_W^2\right]}
+\frac{4m_W^2}{(p^2-m_W^2)\left[(p-k_1)^2-m_W^2\right]}
\right.\nonumber\\&\left.
-\frac{\frac{1}{2}m_H^2+7m_W^2}{(p^2-m_W^2)\left[(p-k_1-k_2)^2-m_W^2\right]}
+\frac{4m_W^2\left[2(1-n)m_W^2-m_H^2\right]}{(p^2-m_W^2)\left[(p-k_1)^2-m_W^2\right]\left[(p-k_1-k_2)^2-m_W^2\right]}
\right]
\\
i{\cal M}_4=&\frac{e^2g}{m_W^3}
\left[
-\frac{1}{2(p^2-m_W^2)}-\frac{1}{2\left[(p-k_1-k_2)^2-m_W^2\right]}
-\frac{4m_W^2}{\left[(p-k_1)^2-m_W^2\right]\left[(p-k_1-k_2)^2-m_W^2\right]}
\right.\nonumber\\&\left.
+\frac{\frac{1}{2}m_H^2+7m_W^2}{(p^2-m_W^2)\left[(p-k_1-k_2)^2-m_W^2\right]}
\right]
\end{flalign}
The integral ${\cal M}_1-{\cal M}_5$ can be expanded using Passarino-Veltman integrals \cite{Passarino:1978jh}.
These integrals can further be reduced to the scalar integrals $A_0$, $B_0$ and $C_0$.
The results are
\begin{eqnarray}
\int\frac{d^np}{(2\pi)^n}
{\cal M}_1 g^{\mu\nu}&=&
\frac{e^2g}{(4\pi)^2m_W}
\left[4m_W^2(1-2m_H^2C_0(m_H^2,0,0,m_W^2,m_W^2,m_W^2))
\right.\nonumber\\&&\left.
-(m_H^2+6m_W^2)B_0(m_H^2,m_W^2,m_W^2)\right]g^{\mu\nu}
\\
\int\frac{d^np}{(2\pi)^n}
{\cal M}_2 p^\mu p^\nu&=&
\frac{e^2g}{(4\pi)^2m_Wm_H^2}
\left[
(m_H^2+6m_W^2)\left(1+B_0(m_H^2,m_W^2,m_W^2)
\right.\right.\nonumber\\&&\left.\left.
+2m_W^2C_0(m_H^2,0,0,m_W^2,m_W^2,m_W^2)\right)(m_H^2g^{\mu\nu}-2k_2^\mu k_1^\nu)
-4m_W^2m_H^2g^{\mu\nu}
\right.\nonumber\\&&\left.
+2(m_H^2+6m_W^2)\left(2B_0(0,m_W^2,m_W^2)-B_0(m_H^2,m_W^2,m_W^2)\right)k_2^\mu k_1^\nu
\right]
\\
\int\frac{d^np}{(2\pi)^n}
{\cal M}_3 p^\mu k_1^\nu&=&
\frac{e^2g}{(4\pi)^24m_W^3m_H^2}
\left[
16m_W^2(m_H^2+6m_W^2)-2(7m_H^2+48m_W^2)A_0(m_W^2)
\right.\nonumber\\&&\left.
+\left(96m_W^4+2m_W^2m_H^2-m_H^4\right)B_0(m_H^2,m_W^2,m_W^2)
\right]k_2^\mu k_1^\nu
\\
\int\frac{d^np}{(2\pi)^n}
{\cal M}_4 k_2^\nu p^\nu&=&
\frac{e^2g}{(4\pi)^24m_W^3}
\left[
16m_W^2-18A_0(m_W^2)
+(m_H^2+14m_W^2)B_0(m_H^2,m_W^2,m_W^2)
\right]k_2^\mu k_1^\nu\nonumber\\
\\
\int\frac{d^np}{(2\pi)^n}
{\cal M}_5 k_2^\mu k_1^\nu&=&
\frac{e^2g}{(4\pi)^2m_W}
\left[4B_0(0,m_W^2,m_W^2)+16m_W^2C_0(m_H^2,0,0,m_W^2,m_W^2,m_W^2)
\right]k_2^\mu k_1^\nu
\nonumber\\
\end{eqnarray}
Using $B_0(0,x,x)=A_0(x)/x-1$, these add up to
\begin{eqnarray}\label{eq:uresult}
{\cal M}&=&\frac{e^2g}{(4\pi)^2}\frac{1}{m_H^2m_W}
\left[m_H^2+6m_W^2-6m_W^2(m_H^2-2m_W^2)C_0(m_H^2,0,0,m_W^2,m_W^2,m_W^2)\right]
\nonumber\\&&\times
\left(m_H^2g^{\mu\nu}-2k_2^\mu k_1^\nu\right)\epsilon_\mu(k_1) \epsilon_\nu(k_2)
\end{eqnarray}
The expression for the $C_0$ function is known to be
\begin{equation}
C_0(m_H^2,0,0,m_W^2,m_W^2,m_W^2)=\frac{-2}{m_H^2}f\left(\frac{4m_W^2}{m_H^2}\right)
\end{equation}
where
\begin{equation}
f(\beta)=\left\{
\begin{array}{ll}
\arcsin^2(\beta^{-\frac{1}{2}}) &\quad \mbox{for}\quad \beta\geq 1\\
-\frac{1}{4}\left[\ln\frac{1+\sqrt{1-\beta}}{1-\sqrt{1-\beta}}-i\pi\right]^2&
\quad \mbox{for}\quad \beta<1
\end{array}\right.\;.
\end{equation}
The final result is
\begin{equation}
{\cal M}=\frac{e^2g}{(4\pi)^2m_W}\left[2+3\beta+3(2\beta-\beta^2)f(\beta)\right]
\left[(k_1\cdot k_2)g^{\mu\nu}-k_2^\mu k_1^\nu\right]\epsilon_\mu(k_1) \epsilon_\nu(k_2)
\end{equation}
where
\begin{equation}
\beta=\frac{4m_W^2}{m_H^2}
\end{equation}

\section{$R_\xi$ Gauge}

In the $R_\xi$ gauge, the number of diagrams increases, because Goldstone bosons and ghosts enter at one loop. We show the diagrams in Fig.~\ref{fig:Rxi}.

To simplify the calculation, we divide the $W$ boson propagator into two parts
\begin{equation}\label{eq:wprop}
\frac{-i}{q^2-m_W^2}\left(g^{\mu\nu}-(1-\xi)\frac{q^\mu q^\nu}{q^2-\xi m_W^2}\right)
=\frac{-i}{q^2-m_W^2}\left(g^{\mu\nu}-\frac{q^\mu q^\nu}{m_W^2}\right)
+\frac{-i}{q^2-\xi m_W^2}\frac{q^\mu q^\nu}{m_W^2}
\end{equation}
The first term on the right-hand side is a propagator in the unitary gauge.
The second term has a $q^2-\xi m_W^2$ in the denominator, and thus can be combined
with Goldstone boson and ghost propagators that appear in other diagrams, to simplify the calculation.

Using this method, the diagrams with $W$ propagators are divided into several parts.
For example, the diagram in Fig.~\ref{fig:Rxi}(a) has 8 pieces.
We denote them by ${\cal M}_{ijk}$ where $i,j,k=1,2$ according to which term on the right-hand side of Eq.~(\ref{eq:wprop}) the $W$-propagator takes.
\begin{equation}
{\cal M}_a={\cal M}_{111}+{\cal M}_{112}+{\cal M}_{121}+{\cal M}_{211}+{\cal M}_{122}+{\cal M}_{212}+{\cal M}_{221}+{\cal M}_{222}
\end{equation}
with
\begin{flalign}
{\cal M}_{111}=&\int\frac{d^np}{(2\pi)^n}2V^{\alpha\beta\gamma\delta\lambda\rho\mu\nu}
\frac{g_{\alpha\gamma}-\frac{p_\alpha p_\gamma}{m_W^2}}{p^2-m_W^2}
\frac{g_{\lambda\rho}-\frac{(p-k_1)_\lambda (p-k_1)_\rho}{m_W^2}}{(p-k_1)^2-m_W^2}
\frac{g_{\delta\beta}-\frac{(p-k_1-k_2)_\delta (p-k_1-k_2)_\beta}{m_W^2}}{(p-k_1-k_2)^2-m_W^2}
\epsilon_\mu(k_1) \epsilon_\nu(k_2)
\\
{\cal M}_{112}=&\int\frac{d^np}{(2\pi)^n}2V^{\alpha\beta\gamma\delta\lambda\rho\mu\nu}
\frac{g_{\alpha\gamma}-\frac{p_\alpha p_\gamma}{m_W^2}}{p^2-m_W^2}
\frac{g_{\lambda\rho}-\frac{(p-k_1)_\lambda (p-k_1)_\rho}{m_W^2}}{(p-k_1)^2-m_W^2}
\frac{\frac{(p-k_1-k_2)_\delta (p-k_1-k_2)_\beta}{m_W^2}}{(p-k_1-k_2)^2-\xi m_W^2}
\epsilon_\mu(k_1) \epsilon_\nu(k_2)
\\
{\cal M}_{121}=&\int\frac{d^np}{(2\pi)^n}2V^{\alpha\beta\gamma\delta\lambda\rho\mu\nu}
\frac{g_{\alpha\gamma}-\frac{p_\alpha p_\gamma}{m_W^2}}{p^2-m_W^2}
\frac{\frac{(p-k_1)_\lambda (p-k_1)_\rho}{m_W^2}}{(p-k_1)^2-\xi m_W^2}
\frac{g_{\delta\beta}-\frac{(p-k_1-k_2)_\delta (p-k_1-k_2)_\beta}{m_W^2}}{(p-k_1-k_2)^2-m_W^2}
\epsilon_\mu(k_1) \epsilon_\nu(k_2)
\\
{\cal M}_{211}=&\int\frac{d^np}{(2\pi)^n}2V^{\alpha\beta\gamma\delta\lambda\rho\mu\nu}
\frac{\frac{p_\alpha p_\gamma}{m_W^2}}{p^2-\xi m_W^2}
\frac{g_{\lambda\rho}-\frac{(p-k_1)_\lambda (p-k_1)_\rho}{m_W^2}}{(p-k_1)^2-m_W^2}
\frac{g_{\delta\beta}-\frac{(p-k_1-k_2)_\delta (p-k_1-k_2)_\beta}{m_W^2}}{(p-k_1-k_2)^2-m_W^2}
\epsilon_\mu(k_1) \epsilon_\nu(k_2)
\\
&\cdots\nonumber
\\
{\cal M}_{222}=&\int\frac{d^np}{(2\pi)^n}2V^{\alpha\beta\gamma\delta\lambda\rho\mu\nu}
\frac{\frac{p_\alpha p_\gamma}{m_W^2}}{p^2-\xi m_W^2}
\frac{\frac{(p-k_1)_\lambda (p-k_1)_\rho}{m_W^2}}{(p-k_1)^2-\xi m_W^2}
\frac{\frac{(p-k_1-k_2)_\delta (p-k_1-k_2)_\beta}{m_W^2}}{(p-k_1-k_2)^2-\xi m_W^2}
\epsilon_\mu(k_1) \epsilon_\nu(k_2)
\end{flalign}
where
\begin{eqnarray}
V^{\alpha\beta\gamma\delta\lambda\rho\mu\nu}
=-ie^2gm_Wg^{\alpha\beta}
\left[
(2p-k_1)^\mu g^{\gamma\lambda}-(p+k_1)^\lambda g^{\mu\gamma}
-(p-2k_1)^\gamma g^{\mu\lambda}
\right]\nonumber\\\times
\left[
-(p-k_1+k_2)^\delta g^{\nu\rho}-(p-k_1-2k_2)^\rho g^{\nu\delta}
+(2p-2k_1-k_2)^\nu g^{\rho\delta}
\right]
\end{eqnarray}
denotes the contribution from the vertices.
A factor of 2 is included to take into account the diagram with the two photons exchanged.
This diagram can be obtained by $k_1\leftrightarrow k_2$ and $\mu\leftrightarrow\nu$.
Since we are only interested in terms that are proportional to either $g^{\mu\nu}$ or $k_2^\mu k_1^\nu$,
the contribution from this diagram is the same.

There are also diagrams with both $W$ and Goldstone boson propagators. We use the same notation,
but with the subscript 0 to denote a Goldstone boson propagator:
\begin{eqnarray}
{\cal M}_c&=&{\cal M}_{110}+{\cal M}_{120}+{\cal M}_{210}+{\cal M}_{220}\\
{\cal M}_e&=&{\cal M}_{100}+{\cal M}_{200}\\
{\cal M}_f&=&{\cal M}_{101}+{\cal M}_{102}+{\cal M}_{201}+{\cal M}_{202}\\
{\cal M}_g&=&{\cal M}_{010}+{\cal M}_{020}\\
{\cal M}_h&=&{\cal M}_{000}
\end{eqnarray}
For ${\cal M}_c$, we have
\begin{flalign}
{\cal M}_{110}=&\int\frac{d^np}{(2\pi)^n}(-4)V'^{\alpha\gamma\lambda\rho\mu\nu}
\frac{g_{\alpha\gamma}-\frac{p_\alpha p_\gamma}{m_W^2}}{p^2-m_W^2}
\frac{g_{\lambda\rho}-\frac{(p-k_1)_\lambda (p-k_1)_\rho}{m_W^2}}{(p-k_1)^2-m_W^2}
\frac{1}{(p-k_1-k_2)^2-\xi m_W^2}
\epsilon_\mu(k_1) \epsilon_\nu(k_2)
\\
{\cal M}_{120}=&\int\frac{d^np}{(2\pi)^n}(-4)V'^{\alpha\gamma\lambda\rho\mu\nu}
\frac{g_{\alpha\gamma}-\frac{p_\alpha p_\gamma}{m_W^2}}{p^2-m_W^2}
\frac{\frac{(p-k_1)_\lambda (p-k_1)_\rho}{m_W^2}}{(p-k_1)^2-\xi m_W^2}
\frac{1}{(p-k_1-k_2)^2-\xi m_W^2}
\epsilon_\mu(k_1) \epsilon_\nu(k_2)
\\
{\cal M}_{210}=&\int\frac{d^np}{(2\pi)^n}(-4)V'^{\alpha\gamma\lambda\rho\mu\nu}
\frac{\frac{p_\alpha p_\gamma}{m_W^2}}{p^2-\xi m_W^2}
\frac{g_{\lambda\rho}-\frac{(p-k_1)_\lambda (p-k_1)_\rho}{m_W^2}}{(p-k_1)^2-m_W^2}
\frac{1}{(p-k_1-k_2)^2-\xi m_W^2}
\epsilon_\mu(k_1) \epsilon_\nu(k_2)
\\
{\cal M}_{220}=&\int\frac{d^np}{(2\pi)^n}(-4)V'^{\alpha\gamma\lambda\rho\mu\nu}
\frac{\frac{p_\alpha p_\gamma}{m_W^2}}{p^2-\xi m_W^2}
\frac{\frac{(p-k_1)_\lambda (p-k_1)_\rho}{m_W^2}}{(p-k_1)^2-\xi m_W^2}
\frac{1}{(p-k_1-k_2)^2-\xi m_W^2}
\epsilon_\mu(k_1) \epsilon_\nu(k_2)
\end{flalign}
and
\begin{equation}
V'^{\alpha\gamma\lambda\rho\mu\nu}=i\frac{1}{2}e^2gm_W(p-2k_1-2k_2)^\alpha
\left[
(2p-k_1)^\mu g^{\gamma\lambda}-(p+k_1)^\lambda g^{\mu\gamma}
-(p-2k_1)^\gamma g^{\mu\lambda}
\right]g^{\nu\rho}
\end{equation}
Similarly for ${\cal M}_{e,f,g,h}$.
These terms all include a factor of 2 from exchanging the external photons.
Diagrams (c) and (e) have another factor of 2, due to contributions from diagrams
with opposite charge in the loop.

Diagrams in Fig.~3 (b, d, i) only have two propagators. We denote them by
\begin{eqnarray}
{\cal M}_b&=&{\cal M}_{11}+{\cal M}_{12}+{\cal M}_{21}+{\cal M}_{22}\\
{\cal M}_d&=&{\cal M}_{10}+{\cal M}_{20}\\
{\cal M}_i&=&{\cal M}_{00}
\end{eqnarray}
The notation is similar to before. For example,
\begin{flalign}
{\cal M}_{11}=&\int\frac{d^np}{(2\pi)^n}ie^2gm_Wg^{\alpha\beta}S^{\mu\nu,\gamma\delta}
\frac{g_{\alpha\gamma}-\frac{p_\alpha p_\gamma}{m_W^2}}{p^2-m_W^2}
\frac{g_{\delta\beta}-\frac{(p-k_1-k_2)_\delta (p-k_1-k_2)_\beta}{m_W^2}}{(p-k_1-k_2)^2-m_W^2}
\epsilon_\mu(k_1) \epsilon_\nu(k_2)
\\
{\cal M}_{12}=&\int\frac{d^np}{(2\pi)^n}ie^2gm_Wg^{\alpha\beta}S^{\mu\nu,\gamma\delta}
\frac{g_{\alpha\gamma}-\frac{p_\alpha p_\gamma}{m_W^2}}{p^2-m_W^2}
\frac{\frac{(p-k_1-k_2)_\delta (p-k_1-k_2)_\beta}{m_W^2}}{(p-k_1-k_2)^2-\xi m_W^2}
\epsilon_\mu(k_1) \epsilon_\nu(k_2)
\\
{\cal M}_{21}=&\int\frac{d^np}{(2\pi)^n}ie^2gm_Wg^{\alpha\beta}S^{\mu\nu,\gamma\delta}
\frac{\frac{p_\alpha p_\gamma}{m_W^2}}{p^2-\xi m_W^2}
\frac{g_{\delta\beta}-\frac{(p-k_1-k_2)_\delta (p-k_1-k_2)_\beta}{m_W^2}}{(p-k_1-k_2)^2-m_W^2}
\epsilon_\mu(k_1) \epsilon_\nu(k_2)
\\
{\cal M}_{22}=&\int\frac{d^np}{(2\pi)^n}ie^2gm_Wg^{\alpha\beta}S^{\mu\nu,\gamma\delta}
\frac{\frac{p_\alpha p_\gamma}{m_W^2}}{p^2-\xi m_W^2}
\frac{\frac{(p-k_1-k_2)_\delta (p-k_1-k_2)_\beta}{m_W^2}}{(p-k_1-k_2)^2-\xi m_W^2}
\epsilon_\mu(k_1) \epsilon_\nu(k_2)
\end{flalign}
and $S_{\mu\nu,\gamma\delta}=2g_{\mu\nu}g_{\gamma\delta}-g_{\mu\gamma}g_{\nu\delta}-g_{\mu\delta}g_{\nu\gamma}$. Similarly for ${\cal M}_d$ and ${\cal M}_i$.

Lastly, there is a ghost loop diagram:
\begin{equation}
{\cal M}_j=\int\frac{d^np}{(2\pi)^n}2ie^2gm_W\xi\frac{(p-k_1)^\mu (p-k_1-k_2)^\nu}
{(p^2-\xi m_W^2)[(p-k_1)^2-\xi m_W^2][(p-k_1-k_2)^2-\xi m_W^2]}
\epsilon_\mu(k_1) \epsilon_\nu(k_2)
\end{equation}
${\cal M}_j$ has a factor of $-1$ from the ghost loop.
Diagrams (d) and (j) contain a factor of 4 from exchanging the external photons and from charge conjugation.

Some of these terms vanish:
\begin{equation}
{\cal M}_{122}={\cal M}_{221}={\cal M}_{222}={\cal M}_{220}=0
\end{equation}

Now we can start to combine these terms. First of all, the sum of ${\cal M}_{111}$ and ${\cal M}_{11}$
should reproduce the full result in Eq.~(\ref{eq:uresult}),
because the first term in the $W$ propagator is
the same as a propagator in the unitary gauge. Since the result must be $\xi$-independent,
we expect all the other terms cancel.

In the remaining terms, certain combinations will give simple results.
For example, the contribution from pure Goldstone boson loops is gauge invariant:
\begin{flalign}
&{\cal M}_{000}+{\cal M}_{00}={\cal M}_h+{\cal M}_i
\nonumber\\
=&-i\frac{e^2gm_H^2}{m_W}\int\frac{d^np}{(2\pi)^n}\left[
\frac{4p^\mu(p-k_1)^\nu}{(p^2-\xi m_W^2)[(p-k_1)^2-\xi m_W^2][(p-k_1-k_2)^2-\xi m_W^2]}\right.\nonumber\\
&\left.-\frac{g^{\mu\nu}}{(p^2-\xi m_W^2)[(p-k_1-k_2)^2-\xi m_W^2]}
\right]\epsilon_\mu(k_1) \epsilon_\nu(k_2)
\nonumber\\
=&\frac{2e^2g}{(4\pi)^2m_W}\left[1+2\xi m_W^2 C_0(m_H^2,0,0,\xi m_W^2,\xi m_W^2,\xi m_W^2)\right][(k_1\cdot k_2)g^{\mu\nu}-k_2^\mu k_1^\nu]\epsilon_\mu(k_1) \epsilon_\nu(k_2)
\end{flalign}
All the remaining terms with no 1 in the subscript should be combined. We find
\begin{flalign}
&{\cal M}_{20}+{\cal M}_{200}+{\cal M}_{202}+{\cal M}_{020}+{\cal M}_j
\nonumber\\=&
i\frac{e^2g}{m_W}\int\frac{d^np}{(2\pi)^n}\left[
\frac{4m_H^2p^\mu(p-k_1)^\nu}{(p^2-\xi m_W^2)[(p-k_1)^2-\xi m_W^2][(p-k_1-k_2)^2-\xi m_W^2]}\right.
\nonumber\\
&\left.+\frac{3p^\mu(p-k_1)^\nu}{[(p-k_1)^2-\xi m_W^2][(p-k_1-k_2)^2-\xi m_W^2]}-\frac{3p^\mu(p-k_1)^\nu}{(p^2-\xi m_W^2)[(p-k_1)^2-\xi m_W^2]}
\right]\epsilon_\mu(k_1) \epsilon_\nu(k_2)
\end{flalign}
The last two terms cancel each other under $p_1\leftrightarrow p_2,\ \mu\leftrightarrow \nu$ and momentum shifting. The first term then gives
\begin{flalign}
&{\cal M}_{20}+{\cal M}_{200}+{\cal M}_{202}+{\cal M}_{020}+{\cal M}_j
\nonumber\\=&
-\frac{e^2g}{(4\pi)^2m_W}\left\{
2\left[1+2\xi m_W^2 C_0(m_H^2,0,0,\xi m_W^2,\xi m_W^2,\xi m_W^2)\right]
\left[(k_1\cdot k_2)g^{\mu\nu}-k_2^\mu k_1^\nu\right]
\right.\nonumber\\
&\quad\quad\left.+m_H^2B_0(m_H^2,\xi m_W^2,\xi m_W^2)g^{\mu\nu}\right\}
\epsilon_\mu(k_1) \epsilon_\nu(k_2)
\end{flalign}
The first term on the right-hand side cancels the contribution
from ${\cal M}_{000}$ and ${\cal M}_{00}$. The second term with a $B_0$ function is cancelled by ${\cal M}_{22}+{\cal M}_{212}+{\cal M}_{210}+{\cal M}_{010}$. In fact,
\begin{flalign}
&{\cal M}_{22}+{\cal M}_{212}+{\cal M}_{210}+{\cal M}_{010}
\nonumber\\
=&-i\frac{e^2g}{m_W^3}\int\frac{d^np}{(2\pi)^n}\left[
\frac{1}{2}\left(p^\mu k_1^\nu-k_2^\mu p^\nu\right)
\left(\frac{1}{p^2-\xi m_W^2}+\frac{1}{(p-k_1-k_2)^2-\xi m_W^2}\right)
\right.\nonumber\\&\left.
-\left(p^\mu k_1^\nu-p^\mu p^\nu\right)\frac{m_W^2}{(p-k_1)-m_W^2}
\left(\frac{1}{p^2-\xi m_W^2}-\frac{1}{(p-k_1-k_2)^2-\xi m_W^2}\right)
\right.\nonumber\\&\left.
+\frac{\left(\xi m_W^2-\frac{1}{2}m_H^2\right)\left(p^\mu k_1^\nu-k_2^\mu p^\nu\right)}
{(p^2-\xi m_W^2)[(p-k_1-k_2)^2-\xi m_W^2]}
-p\cdot(k_1-k_2)g^{\mu\nu}\left(\frac{1}{p^2-\xi m_W^2}+\frac{1}{(p-k_1-k_2)^2-\xi m_W^2}\right)
\right.\nonumber\\&\left.
-\left((1-\xi)m_W^2+\frac{1}{2}m_H^2\right)g^{\mu\nu}\left(\frac{1}{p^2-\xi m_W^2}-\frac{1}{(p-k_1-k_2)^2-\xi m_W^2}\right)
\right.\nonumber\\&\left.
-\frac{m_W^4g^{\mu\nu}}{(p-k_1)-m_W^2}\left(\frac{1}{p^2-\xi m_W^2}-\frac{1}{(p-k_1-k_2)^2-\xi m_W^2}\right)
\right.\nonumber\\&\left.
+\frac{m_H^2\left((1-\xi)m_W^2+\frac{1}{2}m_H^2\right)+(4\xi m_W^2-2m_H^2)p\cdot k_2}
{(p^2-\xi m_W^2)[(p-k_1-k_2)^2-\xi m_W^2]}g^{\mu\nu}
\right]\epsilon_\mu(k_1) \epsilon_\nu(k_2)
\end{flalign}
It's not hard to see that under $k_1\leftrightarrow k_2,\ \mu\leftrightarrow \nu$ and momentum shifting, all terms except the last term cancel out. We have
\begin{equation}
{\cal M}_{22}+{\cal M}_{212}+{\cal M}_{210}+{\cal M}_{010}=\frac{e^2g}{(4\pi)^2m_W}m_H^2B_0(m_H^2,\xi m_W^2,\xi m_W^2)g^{\mu\nu}\epsilon_\mu(k_1) \epsilon_\nu(k_2)
\end{equation}

All the remaining ${\cal M}'s$ should cancel. We find
\begin{equation}
{\cal M}_{12}+{\cal M}_{21}+{\cal M}_{112}+{\cal M}_{211}+{\cal M}_{110}+{\cal M}_{10}=-({\cal M}_{121}+{\cal M}_{101})
\end{equation}
and
\begin{equation}
{\cal M}_{120}+{\cal M}_{100}=-2{\cal M}_{102}=-2{\cal M}_{201}
\end{equation}
These all add up to zero, as expected. Thus we see that all terms except ${\cal M}_{11}+{\cal M}_{111}$ are cancelled.
We then obtain the same result as in Eq.~(\ref{eq:uresult}).


\end{document}